\documentclass[%
 reprint,
 amsmath,amssymb,
 aps,
]{revtex4-2}

\usepackage{graphicx}
\usepackage{dcolumn}
\usepackage{bm}
\usepackage{xcolor}
\usepackage{url}
\usepackage[utf8]{inputenc}

\begin{document}


\title{Proposal of a consistent value for the mean excitation energy of liquid water using advanced modeling, detailed simulations and precission proton Bragg curves
}

\author{Pablo de Vera\textsuperscript{1}}
\author{Isabel Abril\textsuperscript{2}}
\author{Flávio Matias\textsuperscript{3}}
\author{Julian M. B. Shorto\textsuperscript{3}}
\author{Hélio Yoriyaz\textsuperscript{3}}
\author{Rafael Garcia-Molina\textsuperscript{1}}

\affiliation{\textsuperscript{1}Departamento de Física, Centro de Investigación en Óptica y Nanofísica, Universidad de Murcia, 30100 Murcia, Spain}

\affiliation{\textsuperscript{2}Departament de Física, Universitat d’Alacant, 03080 Alacant, Spain}

\affiliation{\textsuperscript{3}Instituto de Pesquisas Energéticas e Nucleares, Av. Professor Lineu Prestes, 05508-000 São Paulo, Brazil}







\date{\today}

\begin{abstract}
Protontherapy precision depends on accurately knowing the stopping power of liquid water, as it is the most abundant component of soft body tissues. 
Difficulties when working with volatile liquids have led to a long-standing debate regarding the values of the stopping power and the mean excitation energy $I$ of liquid water. The equivalence between the stopping power of liquid water and amorphous ice per unit mass density across all clinically relevant proton energies opens the possibility to obtain a reliable estimation of the $I$-value of liquid water. In this work, we employ two complementary theoretical methodologies to assess the proton stopping power of water. The MELF-GOS (Mermin Energy Loss Function – Generalized Oscillator Strengths) approach, based on the perturbative dielectric formalism, serves as an accurate reference for sufficiently high energies, while the novel non-perturbative TDDFT-Penn (time-dependent density functional theory – Penn) method ensures accuracy for energies even below the stopping maximum. 
These stopping-power values were incorporated into the simulation code SEICS (Simulation of Energetic Ions and Clusters through Solids) to evaluate proton Bragg curves from $1$ to $230$ MeV. Comparisons with available precise depth-dose measurements provide critical insights that help establish the consistent value $I = 79.4$ eV for liquid water, providing robust physical grounds for improving range prediction for treatment planning. 


\end{abstract}

\maketitle


\section{\label{sec:intro}Introduction}

In the last decades, there has been a significant and constant increase in the number of protontherapy centers being commissioned worldwide, as well as in the number of treated patients \cite{Georgieva2024,PTCOG}. Cancer treatment with energetic proton beams offers multiple advantages over conventional radiotherapy, including selective dose delivery to the tumor region while sparing surrounding healthy tissues, as well as superior relative biological effectiveness \cite{Schardt2010}. Protontherapy benefits from the characteristic shape of its depth-dose curve, known as the Bragg peak: a sharp maximum in the energy deposited appears towards the end of the projectiles' trajectories in tissue, allowing for the precise dose delivery exactly where it is needed.

Still, its optimal application requires exquisite positioning of the Bragg peak inside the patient, as any small deviation (even millimetric) may result in high doses to healthy tissues that are not intended to be irradiated. The range at which the protons can penetrate is mainly determined by the stopping power of the biological medium, i.e., the mean energy lost by the proton per unit distance traveled. As a consequence, there has been great interest in precisely determining either the penetration range of protons in liquid water (the main constituent of soft tissues) by direct measurement \cite{Bichsel2000,Kramer2000,Kumazaki2007,Schardt2008,Faddegon2015} or by means of an accurate estimation of its stopping power \cite{Shimizu2009,Shimizu2010,Siiskonen2011}. However, the fact that water is a volatile liquid makes the direct experimental determination of its stopping power extremely challenging  \cite{Shimizu2009,Shimizu2010,GarciaMolina2013NIMB,deVera2019jets}. On the contrary, measurements on amorphous ice are more straightforward \cite{Kamitsubo1974,Bauer1994}, but the question arises as to how equivalent these two phases of water are.

Related to this discussion is the determination of the mean excitation energy $I$ of liquid water, which is the main parameter defining the target that enters the Bethe equation for the calculation of the stopping power of high-energy charged particles \cite{salvat2022}. In 1985, the International Commission on Radiation Units and Measurements (ICRU) recommended $I = 75$ eV for liquid water \cite{ICRU37}. Even though there has been accumulating evidence since then of the $I$-value being larger (around $78$ - $83$ eV) \cite{Emfietzoglou2009,GarciaMolina2009,Sigmund2009,Paul2013ACQ,Dingfelder2014,Faddegon2015}, to date, important databases such as PSTAR \cite{Berger2005} from the National Institute of Standards and Technology (NIST) still maintain $I = 75$ eV for liquid water.

From a theoretical point of view, the dielectric formalism nowadays allows for reliable calculations of charged-particle stopping powers and $I$-values for condensed-phase materials \cite{Abril1998,HerediaAvalos2005a,GarciaMolina2009,Emfietzoglou2009,Dingfelder2014,deVera2023}. However, this approach, being of perturbative character, is not very accurate for proton energies below $\sim 200$ keV. Recently, the non-perturbative time-dependent density functional theory-Penn (TDDFT-Penn) approach was proposed to overcome such a limitation \cite{Matias2024,Matias2025}. This method enabled excellent reproduction of the experimental proton stopping powers for water vapor, amorphous ice, and liquid water across a wide energy range \cite{Matias2025}. Strikingly, these calculations demonstrated that the mass stopping power (i.e., stopping power per unit mass density) of liquid water and amorphous ice is equivalent over the entire energy range. These methods predict a mean excitation energy of liquid water of $79.4$ eV \cite{GarciaMolina2009,Matias2025}.

By means of Monte Carlo simulations performed with the SEICS (Simulation of Energetic Ions and Clusters through Solids) code \cite{GarciaMolina2011,deVera2018}, in the present work, we will study how sensitive the determination of the proton depth-dose curves is to the uncertainty in the stopping power of liquid water, namely, how simulations are affected by feeding them the results of the dielectric formalism \cite{GarciaMolina2009}, the TDDFT-Penn approach \cite{Matias2025}, or simply by interpolation and extrapolation of currently available experimental data \cite{Shimizu2009,Shimizu2010,Siiskonen2011}. By doing so, it will be assessed to what extent the details of the stopping power at energies around and below the energy-loss maximum are relevant for practical applications in protontherapy. Importantly, current simulations will also serve to evaluate the impact of the $I$-value on the determination of proton depth-dose curves and will provide solid physical grounds for the precise estimation of the mean excitation energy of liquid water.

The manuscript is structured as follows: in Section \ref{sec:methods} we review the dielectric formalism (section \ref{sec:MELF-GOS}) and TDDFT-Penn (section \ref{sec:TDDFT-Penn}) methods, as well as the currently available experimental determinations for the proton stopping power of liquid water (section \ref{sec:jets}). The fundamentals of the SEICS code will also be reviewed (section \ref{sec:SEICS}). In Section \ref{sec:stopping}, the stopping power of liquid water and amorphous ice will be discussed in light of the presented methods, and a review of the $I$-values proposed for liquid water will be offered. In section \ref{sec:lowE}, the results of depth-dose curve simulations for low-energy proton beams will be analyzed in order to discuss the effect of the details of the stopping power on their assessment. Then, simulations for proton beams at clinical energies will be discussed in section \ref{sec:highE}, where important insights into the $I$-value of liquid water will be obtained. Final conclusions are drawn in section \ref{sec:conclusions}, and a recommendation for the mean excitation energy of liquid water will be given based on the results of the theoretical models and the simulations.

\section{\label{sec:methods}Materials and methods}

\subsection{\label{sec:MELF-GOS}Perturbative calculation of proton stopping power: the dielectric formalism and the MELF-GOS method}

The dielectric formalism and the MELF-GOS (Mermin Energy Loss Function -- Generalized Oscillator Strengths) method \cite{Abril1998,HerediaAvalos2005a,GarciaMolina2009} constitute a reliable approach for calculating stopping quantities of charged particle beams in condensed matter, showing excellent agreement with a large collection of experimental determinations in many materials \cite{deVera2011a,deVera2014b,deVera2023}. However, uncertainties remain around the stopping-power maximum and below ($\leq 100-200$ keV/u), as this method is perturbative and based on the first Born approximation. 

Within the dielectric framework, the stopping power (mean energy loss per unit path length traveled by the projectile) of a proton of mass $M$, atomic number $Z$, and charge state $q$, with kinetic energy $T$, is calculated as:
\begin{equation}
    S_q (T) = \frac{e^2}{\pi \hbar^2} \frac{M}{T} \int_{E_-}^{E_+}E\, {\rm d}E \int_{k_{-}}^{k_{+}}  \frac{{\rm d}k}{k} [Z-\rho_q(k)]^2 \, {\rm Im}\left[ \frac{-1}{\varepsilon(k,E)} \right] \, \mbox{,}
    \label{eq:Sq}
\end{equation}
where $e$ is the elementary charge, $\hbar$ is the reduced Planck's constant, and $\rho_q(k)$ is the Fourier's transform of the projectile's electronic density. $E$ is the energy transferred in an inelastic collision, while $\hbar k$ is the momentum transfer. The quantity ${\rm Im}\left[ \frac{-1}{\varepsilon(k,E)} \right]$ is known as the Energy Loss Function (ELF), which represents the electronic excitation spectrum of the target and depends on its complex dielectric function $\varepsilon(k,E)$. The second momentum of the energy loss distribution, $\Omega^2_q (T)$, known as energy loss straggling, can be obtained by a similar equation, but replacing $E$ with $E^2$ in the energy integral.
The integration limits $E_+$ and $E_-$, as well as $k_+$ and $k_-$, are obtained by the fulfillment of the laws of conservation of energy and momentum \cite{Abril1998, deVera2023}.

The ELF can be divided in excitations of the valence (i.e., outher-shell) and the inner-shell electrons of the target \cite{Abril1998,HerediaAvalos2005a}:
\begin{equation}
    {\rm Im}\left[ \frac{-1}{\varepsilon(k,E)} \right] = {\rm Im}\left[ \frac{-1}{\varepsilon(k,E)} \right]_{\mathrm {out}} + \sum_j \nu_j \sum_{nl} {\rm Im}\left[ \frac{-1}{\varepsilon(k,E)} \right]_{nl}^{j} \, \mbox{.}
    \label{eq:ELF}
\end{equation}
The first term corresponds to the excitation of the valence electrons, while in the second term, $\nu_j$ refers to the stoichiometric coefficient of the atomic constituent $j$ of the target. Inner-shell excitations are summed over all the possible combinations of the atomic quantum numbers $n$ and $l$. These contributions can be calculated by means of hydrogenic generalized oscillator strengths (GOS), $\frac{{\rm d}f_{nl}^j(k, E)}{{\rm d} E}$, as \cite{Egerton2011}:
\begin{equation}
{\rm Im}\left[ \frac{-1}{\epsilon(k,E)}\right]_{nl}^{j} = \frac{2\pi^2 \hbar^2 e^2 {\cal N}}{m E} \,
\frac{{\rm d}f_{nl}^j(k, E)}{{\rm d} E} \, \Theta(E - E_{{\rm th},nl}^{j})\, \mbox{,}
\label{GOS}
\end{equation}
where ${\cal N}$ is the molecular density and $m$ is the electron mass. $\Theta(E - E_{{\rm th},nl}^{j})$ is the step function, preventing excitations for energy transfers below the inner shell binding energy $E_{{\rm th},nl}^{j}$.

The ELF corresponding to the valence electrons can be defined, in the optical limit ($k=0$), as a weighted sum of Drude-type functions ${\rm Im}\left[ \frac{-1}{\epsilon(E_i,\gamma_i;k=0,E)} \right]_{i}$:
\begin{equation}
{\rm Im}\left[ \frac{-1}{\epsilon(k=0,E)} \right]_{\rm out} = \sum_i \frac{A_i}{E_i^2} {\rm Im}\left[ \frac{-1}{\epsilon(E_i,\gamma_i;k=0,E)} \right]_{i} \nonumber
\end{equation}
\begin{equation}
 = \sum_i \Theta(E - E_{{\rm th},i}) \, \frac{A_i \, E \, \gamma_i}{(E^2-E_i^2)^2+(E \gamma_i)^2} \, \mbox{,} 
 \label{eq:ELFout}
\end{equation}
where $E_i$, $\gamma_i$ and $A_i$ are fitting parameters, corresponding to the energy, width and intensity of the different features of the optical ELF. Again, $\Theta(E - E_{{\rm th},i})$ is a step function allowing only excitations with energy transfer above the electronic excitation threshold.

The valence ELF is commonly measured experimentally \cite{Hayashi2000} (even though it can also be calculated \textit{ab initio} \cite{Taioli2021JPCL}) and then can be fitted at the optical limit by means of Eq. (\ref{eq:ELFout}), ensuring the fullfilment of different sum rules \cite{Tanuma1993a}. The properties of the Mermin dielectric function \cite{Mermin1970} (which, in the optical limit, coincides with the Drude-type function ${\rm Im}\left[ \frac{-1}{\epsilon(E_i,\gamma_i;k=0, E)} \right]_{i}$) allow to calculate the ELF over the entire $E$-$k$ plane. For the case of liquid water, a target for which an extensive set of measurements of the ELF exist over both $E$ and $k$ \cite{Hayashi2000}, the MELF-GOS method shows excellent agreement with experiment \cite{GarciaMolina2009}. The fitting parameters can be found in Ref. \cite{GarciaMolina2009}.

The total stopping power for the projectile is a sum over the contribution of each charge state $q$ (for protons, $q = 1$ or $q=0$), wheighted by the corresponding energy-dependent charge fractions $\phi_q(T)$. The effect of the projectile's electronic cloud polarization, as well as that of electron capture and loss processes, are also accounted for: 
\begin{equation}
S(T) = \sum_q \phi_q(T) S_{{\rm pol},q}(T) + S_{\rm C\&L} \, \mbox{,}
\label{eq:Sions}
\end{equation}
where $S_{{\rm pol},q}(T)$ denotes the stopping power for each charge state, Eq. (\ref{eq:Sq}), after inclusion of the polarization contribution \cite{HerediaAvalos2002a}, while $S_{\rm C\&L}$ corresponds to the capture and loss contribution to the stopping power \cite{Denton2008c}. The charge-fraction values can be obtained from the parameterization provided in the CasP code \cite{Schiwietz2001}.

\subsection{\label{sec:TDDFT-Penn}Non-perturbative calculation of proton stopping power: the TDDFT-Penn approach}

Recently, the TDDFT-Penn (time-dependent density functional theory-Penn) method has been introduced in Refs. \cite{Matias2024,Matias2025}. It is also based on the dielectric properties of the material measured at the optical limit, but it introduces a non-perturbative scheme that accurately calculates the stopping power over an arbitrary energy range, including low proton energies below the stopping maximum. This method has demonstrated high accuracy in determining the stopping power of protons across different phases of water \cite{Matias2025} (as well as other materials, such as polymers \cite{Matias2024}), yielding excellent agreement with experimental measurements for water vapor, amorphous ice, and liquid water.

First, the stopping power is calculated by means of TDDFT \cite{Auth1998,Borisov2004,Borisov2004b,quijada2007,Koval2017} for a proton of a given velocity $v$ crossing the diameter of a nearly constant electron density (jellium) sphere of a given density $n_\mathrm{0} = E_\mathrm{p}^2/(4\pi \hbar^2)$, $E_\mathrm{p}$ being its characteristic plasmon energy:
\begin{equation}
    S_{\text{TDDFT}}(v,E_\mathrm{p}) = \frac{E_{\rm{loss}}(v,E_{\rm p})}{2R_{\text{cl}}} \, \mbox{.}
\label{stp_tddft}
\end{equation}
Here, $E_{\rm{loss}}(v,E_{\rm p})$ is the energy lost by the proton while crossing the jellium, while $R_{\text{cl}}$ is the sphere radius.

Then, the inhomogeneous electron density of the complex target material can be taken into account by integrating over an ensemble of jellium spheres of different densities \cite{Matias2024,Matias2025}:
\begin{equation}
    S_{\text{TDDFT--Penn}}(v) = \int_0^{\infty}{\rm d}E_\mathrm{p} \, g(E_\mathrm{p}) \, 
        S_{\text{TDDFT}}(v,E_\mathrm{p}) \, \mbox{,}
\end{equation}
where the weighting function $g(E_\mathrm{p})$ depends on the optical ELF of the material, as deduced from the Penn method \cite{Vos2019}:
\begin{equation}
    g(E_\mathrm{p}) =\frac{2 \hbar}{\pi E_\mathrm{p}} \, {\rm Im}\left[ \frac{-1}{\varepsilon(k=0,E_\mathrm{p})} \right] \, \mbox{.}
\end{equation}
The optical ELF is defined by Eq. (\ref{eq:ELF}), and its parameters for liquid water are the same as those used within the MELF-GOS method \cite{GarciaMolina2009}.

\subsection{\label{sec:jets}Stopping power based on measurements for liquid water and semiempirical approaches}

In principle, it is also possible to obtain an empirical approximation of the stopping power of liquid water for protons, using experimental data and semiempirical approaches such as the SRIM code \cite{SRIM2013}.

There exist two sources for the experimental stopping power of liquid water for protons: a direct transmission measurement through water foils performed by Siiskonen \textit{et al.} above $4$ MeV \cite{Siiskonen2011} and experiments based on proton transmission through water jets by Shimizu \textit{et al.} \cite{Shimizu2009,Shimizu2010} in the range $0.3-2$ MeV. The latter were analyzed using Monte Carlo simulations, in which the SRIM stopping power \cite{SRIM2013} was scaled until the simulated and measured energy distributions coincided. This approach yields an experimental stopping power roughly $\sim 10$\% lower than the previously mentioned theoretical methods.

Following ICRU Report 49 \cite{ICRU49}, experimental mass stopping power ($s= S/\rho$, where $\rho$ is the mass density) data can be conveniently fitted by the equation proposed by Varelas and Biersak \cite{Varelas1970,AndersenZiegler1977}:
\begin{equation}
    s(T_s) = \frac{s_{\rm low}(T_s)  \, s_{\rm high}(T_s) }{s_{\rm low}(T_s)  + s_{\rm high}(T_s) } \, \mbox{,}
    \label{eq:VarelasBiersak}
\end{equation}
which combines an expression valid for low proton energies:
\begin{equation}
    s_{\rm low}(T_s)  = a_1 T_s^{0.45} \, \mbox{,}
\end{equation}
with another one appropriate for high proton energies:
\begin{equation}
    s_{\rm high}(T_s)  = \frac{a_2}{T_s}\ln{\left( 1+\frac{a_3}{T_s} + a_4 T_s \right)} \, \mbox{.}
\end{equation}
The quantity $T_s = T/(M/u)$ is a scaled kinetic energy of the proton, in which the energy $T$ (in MeV) is divided by the ratio of the proton mass to the atomic mass unit, $M/u = 1.0073$; $a_1$, $a_2$, $a_3$, and $a_4$ are fitting parameters. 

Unfortunately, there are no experimental data on protons in liquid water at energies below $0.3$ MeV. In order to obtain a curve over a wide proton energy range, we extrapolated Shimizu \textit{et al.}'s data \cite{Shimizu2009,Shimizu2010} below $0.3$ MeV by means of SRIM values \cite{SRIM2013}. As the water-jet determinations are based on the scaling of SRIM calculations, SRIM provides a somewhat smooth transition at lower proton energies.

Combining the SRIM values \cite{SRIM2013} below $0.3$ MeV, Shimizu \textit{et al.}'s determinations in the range $0.3-2$ MeV \cite{Shimizu2009,Shimizu2010}, and Siiskonen \textit{et al.}'s measurements in the range $4-15$ MeV \cite{Siiskonen2011}, we have fitted Eq.~(\ref{eq:VarelasBiersak}) to obtain the fitting parameters $a_1 = 3535,5$, $a_2 = 95.1$, $a_3 = 1.1$, and $a_4 = 11.2$, with their units appropriate to obtain $s$ in MeV cm$^2$/g with $T$ in MeV.

\subsection{\label{sec:SEICS}Simulation of Energetic Ions and Clusters through Solids (SEICS) code}

To simulate proton propagation through liquid water and obtain depth-dose curves, we have used the SEICS code \cite{GarciaMolina2011,deVera2018}. SEICS numerically solves the equation of motion of the projectile by updating its coordinates $\vec{r}$ and velocity $\vec{v}$ at time step $i+1$ from the values at the previous time step $i$ by means of the velocity variant of Verlet's algorithm \cite{Allen1989}:
\begin{eqnarray}
\vec{r}_{i+1} & = & \vec{r}_i + \vec{v}_i \Delta t+\frac{\vec{F}_i}{2M}\mbox{ }(\Delta t)^2 \mbox{ }\left[ 1-\left(\frac{v_i}{c}\right)^2 \right]^{3/2} \mbox{ }{\rm ,} \, \label{eq:r-Verlet-rel} \\
\vec{v}_{i+1} & = & \vec{v}_i+\frac{\vec{F}_i+\vec{F}_{i+1}}{2M}\mbox{ }\Delta t \mbox{ }\left[1-\left(\frac{v_i}{c}\right)^2 \right]^{3/2} \mbox{ }{\rm .}
\label{eq:v-Verlet-rel}
\end{eqnarray}
The term in brackets is a correction to account for relativistic speeds \cite{deVera2018}, with $c$ being the speed of light and $\Delta t$ being the simulation time step. The electronic stopping force felt by the projectile is:
\begin{equation}
\vec{F} = -\left[ S_q + \frac{\Omega_q}{\sqrt{\Delta s}} \sqrt{-2\ln{R_1}}\cos{(2 \pi R_2)} \right]  \hat{v} \mbox{
}{\rm ,} \label{eq:A6}
\end{equation}
where $\hat{v}$ is the velocity unit vector, $R_1$ and $R_2$ are random numbers uniformly distributed between $0$ and $1$, and $\Delta s$ is the distance traveled between two consecutive  time steps. The force is drawn from a Gaussian distribution centered at the stopping power $S_q$ and with a variance related to the energy loss straggling, according to the Box-Müller algorithm \cite{Box1958}.

The SEICS code accounts for all physical phenomena relevant to proton transport, as explained elsewhere, including elastic scattering with target nuclei, electron capture and loss processes \cite{GarciaMolina2011}, and nuclear fragmentation reactions with target atoms \cite{deVera2018}.

\section{\label{sec:results}Results and discussion}

\subsection{\label{sec:stopping}The stopping power of liquid water for protons and the $I$-value}

\begin{figure}
    \centering
    \includegraphics[width=0.48\textwidth]{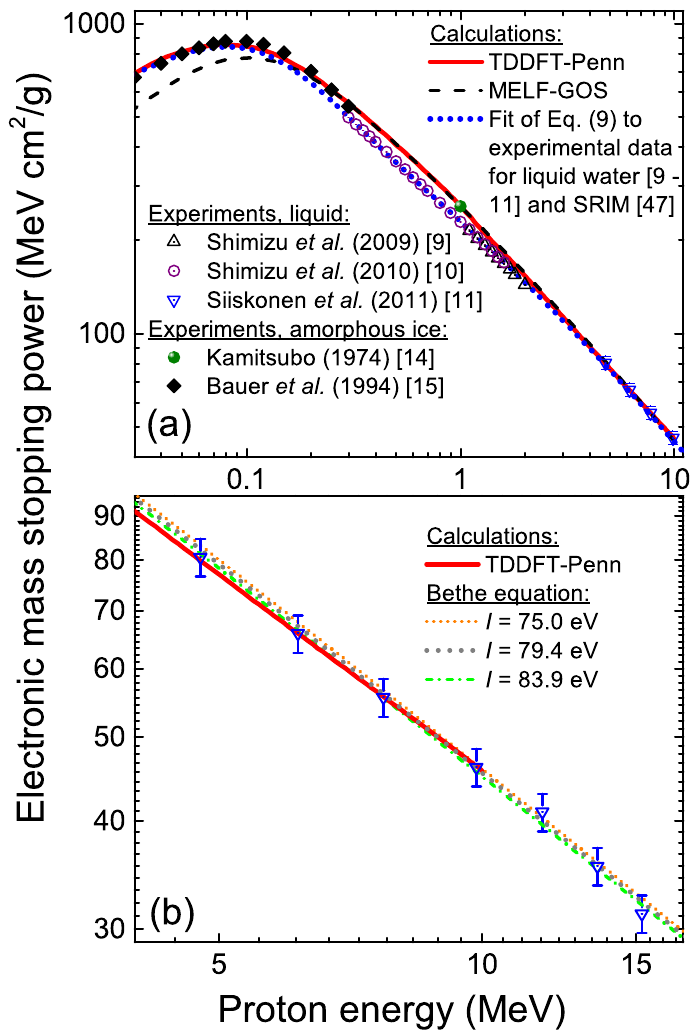}
    \caption{Mass stopping power of water for protons. (a) Calculations for liquid water (lines) are compared to experimental data (open symbols: liquid water; full symbols: amorphous ice) over a wide energy range. (b) Zoom around transmission measurements in liquid water at large energies, together with results of the Bethe equation for different $I$-values.}
    \label{fig:Smodels}
\end{figure}

The three methods reported in sections \ref{sec:MELF-GOS}, \ref{sec:TDDFT-Penn}, and \ref{sec:jets} have been used to compute the mass stopping power of liquid water for protons; the results are 
being plotted in Fig.~\ref{fig:Smodels}. In panel (a), the calculations using the MELF-GOS method are shown by a black dashed line, those of the TDDFT-Penn approach by a solid red line, and the fitting of the Varelas-Biersak equation (\ref{eq:VarelasBiersak}) to the experimental data for liquid water by a dotted blue line. These measured data \cite{Shimizu2009,Shimizu2010,Siiskonen2011} are depicted by open symbols, while full symbols represent the available measurements for amorphous ice \cite{Kamitsubo1974,Bauer1994}.

It should be noted that the MELF-GOS and TDDFT-Penn results converge perfectly for proton energies above $200$ keV. Since dielectric formalism calculations are better suited for large energies, this convergence reinforces the accuracy of both perturbative and non-perturbative models for large energies. The departure of MELF-GOS results below 200 keV may be attributed to the breakdown of the first Born approximation. 

In a previous work \cite{Matias2025}, in which the stopping power of different phases of water (vapor, liquid, hexagonal, and amorphous ices) was studied, it was shown that the mass stopping power of both liquid water and amorphous ice is identical. This can be appreciated from Fig.~\ref{fig:Smodels}(a), as the experimental measurements for amorphous ice \cite{Kamitsubo1974,Bauer1994} perfectly overlap with the TDDFT-Penn calculations over the entire energy range. Shimizu \textit{et al.}'s data for liquid water at intermediate energies \cite{Shimizu2009,Shimizu2010} underestimate our results, while Siiskonen \textit{et al.}'s measurements for liquid water at high energies \cite{Siiskonen2011} perfectly agree with the present calculations. The comparison between these experimental data and TDDFT-Penn calculations can be better seen in Fig.~\ref{fig:Smodels}(b), where the scale is focused around these measurements. 

The fact that Shimizu \textit{et al.}'s \cite{Shimizu2009,Shimizu2010} and Siiskonen \textit{et al.}'s data \cite{Siiskonen2011} (complemented with SRIM \cite{SRIM2013} below $300$ keV; see section \ref{sec:jets}) can be fitted by the Varelas-Biersak formula indicates that, a priori, Shimizu's \textit{et al.}'s experimental determinations, even if lower than other results, may be reasonable. We will denote this curve as the ``Shimizu-Siiskonen-SRIM model'' and use it, together with MELF-GOS and TDDFT-Penn calculations, to study in the next sections the impact of stopping-power uncertainties on simulated Bragg curves. 

For proton energies above approximately $10$ MeV, it is possible to accurately calculate the stopping power by means of the Bethe equation \cite{salvat2022}, which relies on the knowledge of the target's parameter known as the mean excitation energy, $I$. The latter can be conveniently obtained from the optical ELF as:
\begin{equation}
\ln I = \frac{\displaystyle \int_{0}^{\infty} {\rm d}E \, E \ln E \, {\rm Im}\left[{\frac{-1}{\varepsilon(k=0,E)}}\right]}
{\displaystyle \int_{0}^{\infty} {\rm d}E \, E  \, {\rm Im}\left[\frac{-1}{\varepsilon(k=0,E)}\right]} \, \mbox{.}
\label{eq:I}
\end{equation}

\begin{table*}
    \centering
    \resizebox{0.8\textwidth}{!}{
    \begin{tabular}{|c|c|c|}
        \hline
        \textbf{Source} & \textbf{$I$ (eV)} & \textbf{Comments} \\ \hline
        ICRU 37 (1985) \cite{ICRU37} & $75.0 \pm 3$ & Compilation of data prior 1984 \\
        ICRU 73 (2005) \cite{ICRU-Report73} & $67.2$ &  \\
        ICRU 49 (1993) \cite{ICRU49} and PSTAR \cite{Berger2005} & $75.0 \pm 3$ & Based on ICRU 37 (1985) \\ \hline
        Bichsel \textit{et al.} (2000) \cite{Bichsel2000} & $79.7 \pm 2$ & Range in measured ionization curves \\
        & $80.0 \pm 1.3$ &  \\
        Krämer \textit{et al.} (2000) \cite{Kramer2000} & $77$ & Range in measured C Bragg curves \\
        Kumazaki \textit{et al.} (2007) \cite{Kumazaki2007} & $78.4 \pm 1$ & Range in measured H Bragg curves \\
        Schardt \textit{et al.} (2008) \cite{Schardt2008} & $78.0$ & Range in measured H and C Bragg curves \\ \hline
        Errata and Addenda for & $78.0$ & Correction of ICRU 73 (2005) based on \\
        ICRU Report 73 (2009) \cite{Sigmund2009} &  & range measurements by Schardt \textit{et al.} (2008) \\ \hline
        Emfietzoglou \textit{et al.} (2009) \cite{Emfietzoglou2009} & $77.8 \pm 1$ & Dielectric formalism \\
        Garcia-Molina \textit{et al.} (2009) \cite{GarciaMolina2009} & $79.4$ & Dielectric formalism \\
        Dingfelder (2014) \cite{Dingfelder2014} & $78.3$ & Dielectric formalism \\ \hline
        Siiskonen \textit{et al.} (2011) \cite{Siiskonen2011}& $83.9 \pm 2.8$ & Bethe eq. fitted to exp. $S$ data \\ \hline
        Paul (2013) \cite{Paul2013ACQ} & $78.5 \pm 5$ & Analysis of Schardt \textit{et al.} (2008) data \\ \hline
        Faddegon \textit{et al.} (2015) \cite{Faddegon2015} & $80$ -- $83$ & Range in measured $67.5$ MeV H Bragg curve \\ \hline
        This work & $79.4$ & Current perturbative and \\
        &  & non-perturbative calculations \\
        \hline
    \end{tabular}
    }
    \caption{Revision of $I$-values proposed for liquid water since 1985.}
    \label{tab:Is}
\end{table*}

Different $I$-values have been proposed for liquid water over the last few decades by several authors using various experimental and theoretical approaches. Table~\ref{tab:Is} summarizes some of these results, based on the revision provided by Helmut Paul in 2013 \cite{Paul2013ACQ}. Up to 1984, the collected data suggested a value of $75 \pm 3$ eV, which was adopted by ICRU in its reports 37 \cite{ICRU37} and 49 \cite{ICRU49}. It should be noted that despite the accumulation of new data since then, some important databases, such as NIST's PSTAR, still use $I = 75.0$ eV for liquid water \cite{Berger2005}. 

Larger values have been proposed since then, mainly around $78-83$ eV, based on ion beam range measurements \cite{Bichsel2000,Kramer2000,Kumazaki2007,Schardt2008} and on the dielectric formalism \cite{Emfietzoglou2009,GarciaMolina2009,Dingfelder2014}. The only exception was ICRU Report 73 \cite{ICRU-Report73}, which proposed a low value of $67.2$ eV, later on corrected and 
recommending a value of $78$ eV \cite{Sigmund2009}. 

The value obtained from the MELF-GOS method, based on the most recent optical dielectric properties of liquid water measured by Hayashi \textit{et al.} \cite{Hayashi2000}, is $79.4$ eV, supported also by the TDDFT-Penn approach, based on this same data. This is close to dielectric calculations by Emfietzoglou \textit{et al.} ($77.8 \pm 1$ eV) \cite{Emfietzoglou2009} and by Dingfelder \textit{et al.} ($78.3$ eV) \cite{Dingfelder2014}. 
Based on a compilation of previous data, Paul recommended, in 2013, a value of $78.5 \pm 5$ eV \cite{Paul2013ACQ}. It should be noted that the high-energy experiments by Siiskonen \textit{et al.} \cite{Siiskonen2011} can be perfectly fitted by the Bethe equation, yielding a somewhat larger value of $83.9 \pm 2.8$ eV. 



Figure \ref{fig:Smodels}(b) shows the results of the Bethe equation using three reasonable values for $I$, namely $75$ eV (PSTAR value \cite{Berger2005}), $79.4$ eV (MELF-GOS and TDDFT-Penn value, this work), and the largest option of $83.9$ eV (direct fit of the Bethe equation to Siiskonen \textit{et al.}'s experiments \cite{Siiskonen2011}). All three curves agree well with Siiskonen \textit{et al.}'s data \cite{Siiskonen2011} for liquid water, quite well with Bauer \textit{et al.}'s \cite{Bauer1994} and Kamitsubo \textit{et al.}'s data \cite{Kamitsubo1974} for amorphous ice, and also overestimate Shimizu \textit{et al.}'s data \cite{Shimizu2009,Shimizu2010} for liquid water.

Even though the differences in the stopping powers calculated by means of the Bethe equation using the different $I$-values seem very small and all curves fall within experimental error bars (see Fig.~\ref{fig:Smodels}(b)), it will be shown in the next sections that this quantity plays a major role in the determination of proton Bragg curves at the high energies used for clinical applications.

\subsection{\label{sec:lowE}Effect of stopping power models on simulated depth-dose curves at low beam energies}

In the following study, we will examine the effect of different stopping power models (namely, MELF-GOS, TDDFT-Penn, and Shimizu-Siiskonen-SRIM, sections \ref{sec:MELF-GOS}, \ref{sec:TDDFT-Penn}, and \ref{sec:jets}) on the simulated proton Bragg curves in liquid water over a wide energy range. The Bragg curves will be simulated using the SEICS code \cite{GarciaMolina2011,deVera2018}, section \ref{sec:SEICS}. SEICS has previously been used to simulate proton depth-dose curves in liquid water using the MELF-GOS stopping power and energy loss straggling, showing very good agreement with measurements \cite{deVera2018}. However, it should be noted that in that work, comparisons were made with measured Bragg curves for proton beams with initial energies that had relatively large uncertainties. Thus, the simulated energies were slightly modified until the simulated and measured ranges coincided. This point will be further discussed in the next section. It should also be noted that the energy loss straggling needed for the simulations (see section \ref{sec:SEICS}) is not available for models other than MELF-GOS. Thus, for each model, the stopping power provided to the SEICS code will be changed, while MELF-GOS energy loss straggling will always be used.

\begin{figure}
    \centering
    \includegraphics[width=0.45\textwidth]{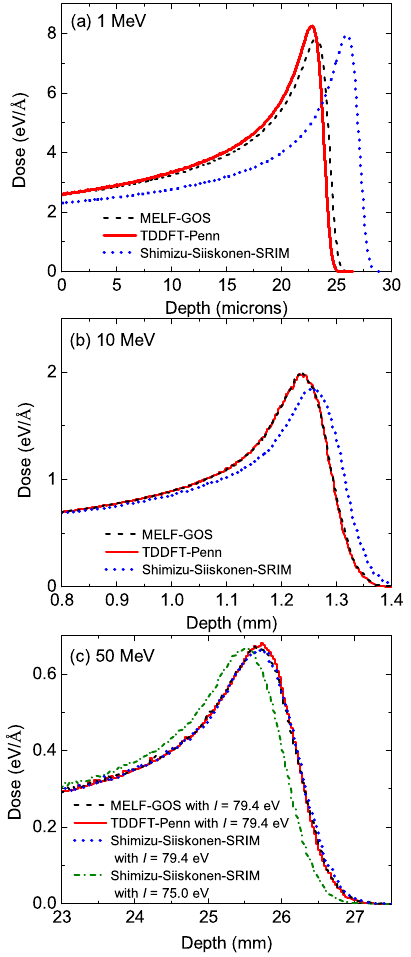}
    \caption{Simulated depth-dose curves for low-energy proton beams of (a) $1$ MeV, (b) $10$ MeV, and (c) $50$ MeV, using different stopping power models. The value of $I = 79.4$ eV is used in all simulations, except for the green dot-dashed curve in panel (c), which uses $I=75.0$ eV.}
    \label{fig:BraggLowT}
\end{figure}

We will start the discussion with low-energy proton beams, for which the effects of each model may be more clearly observed. Higher energies typical of clinical applications will be discussed in the next section. Figure \ref{fig:BraggLowT} shows the depth-dose curves for 1, 10, and 50 MeV protons, simulated using the different stopping power models. 

For 1 MeV, Fig.~\ref{fig:BraggLowT}(a), the results of the MELF-GOS and TDDFT-Penn approaches are rather similar, with a shift in the range (defined as the depth at which the dose falls $80$ \% after the maximum) of just $0.5$ $\mu$m ($\sim 2$ \% difference). Since the Shimizu-Siiskonen-SRIM curve provides somewhat lower stopping powers, its depth-dose curve reaches deeper into the target, with a shift of $3.2$ $\mu$m with respect to the TDDFT-Penn result ($13.5$ \% difference).

The differences are reduced for a larger energy of $10$ MeV, as shown in Fig.~\ref{fig:BraggLowT}(b). For this beam energy, the results given by the TDDFT-Penn and MELF-GOS models are almost indistinguishable. Again, the simulation using the Shimizu-Siiskonen-SRIM data arrives at larger depths, with a shift in the range of $25$ $\mu$m ($\sim 2$ \% difference).


The differences may become insignificant at higher clinical beam energies. Let us study a lower energy that may be used in therapy, e.g., 50 MeV, which could be characteristic for eye tumor treatment. This case is shown in Fig. ~\ref{fig:BraggLowT}(c), where all three depth-dose curves are practically indistinguishable. It should be noted that the contribution from the high-energy stopping power (calculated in the SEICS code using the Bethe equation for energies above $10$ MeV) is very important to the simulation. All three simulations have been performed using the value of $I = 79.4$ eV obtained from both the MELF-GOS and the TDDFT-Penn approaches. However, the Shimizu-Siiskonen-SRIM curve does not imply any specific value of $I$.  To study the influence of the choice of $I$, we show in this figure an additional simulation using $I = 75$ eV, as given by PSTAR \cite{Berger2005}. Clearly, for clinically relevant energies, the mean excitation energy of liquid water will determine the position of the Bragg curve, rather than the details of the low-energy stopping power.


\subsection{\label{sec:highE}Effect of stopping power models on simulated depth-dose curves at clinical beam energies}

As seen from the previous section, for the high energy proton beams characteristic of proton therapy, the differences in the low energy stopping power provided by the various models do not have a large impact on the simulation of depth-dose curves (the differences in cross sections will still be important for other purposes, anyway, e.g., to determine distributions of secondary electrons and their interactions at the micro- and nanoscales, which determine the biological effects of radiation \cite{Solovyov2017}). However, the $I$-value will be important. Both the MELF-GOS and TDDFT-Penn approaches support $I = 79.4$ eV among the suggested options; see Table~\ref{tab:Is}. Thus, in the following, the effect of the choice of $I$ on the simulation of proton depth-dose curves in liquid water at therapeutic energies will be studied. 

\begin{figure*}
    \centering
    \includegraphics[width=0.9\textwidth]{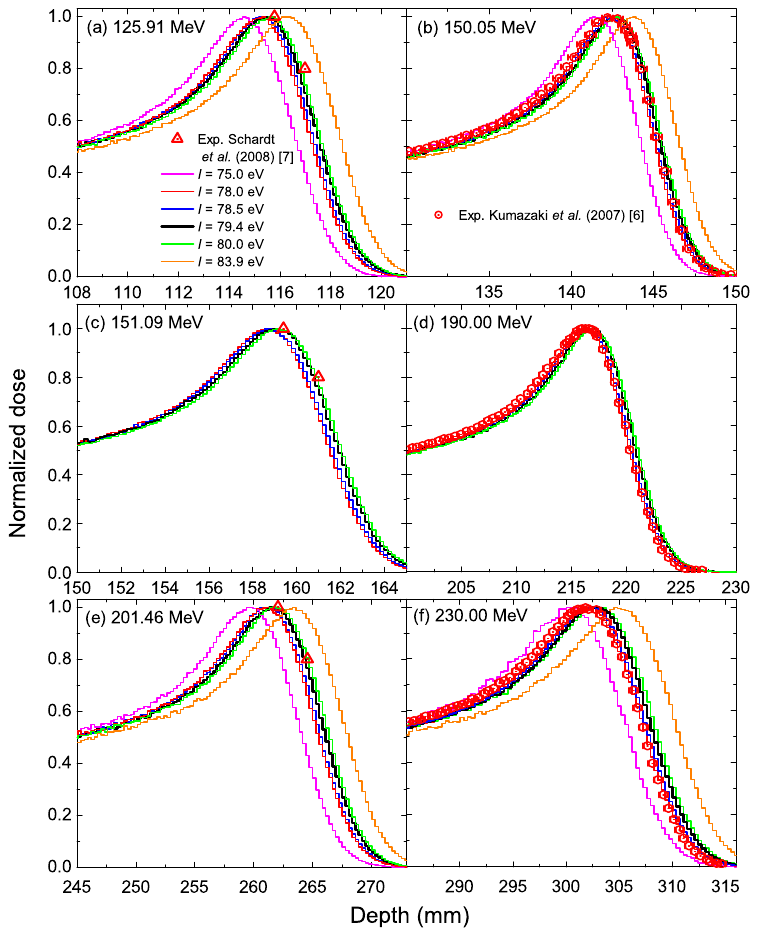}
    \caption{Comparison of simulated (lines) and measured (symbols) depth-dose curves for proton beams of different therapeutic energies. Lines depict simulations using different $I$-values. The two open triangles in the left panels represent the positions of the dose maximum and the range (defined as $80$\% of the dose maximum at the distal part), respectively, as measured by Schardt \textit{et al.} \cite{Schardt2008}. Error bars in open circles represent the uncertainty in the depths measured by Kumazaki \textit{et al.} \cite{Kumazaki2007}.}
    \label{fig:BraggHighT}
\end{figure*}

As noted above, in our previous work simulating Bragg curves \cite{deVera2018}, the initial beam energies were fine-tuned until the simulated and measured ranges coincided, given the uncertainties in the nominal beam energies in experiments. For the present discussion, we need to rely on experiments performed with very well-defined beam energies, so there is no need for this fine-tuning in the simulations. This was achieved in the works by Kumazaki \textit{et al.} \cite{Kumazaki2007} and by Schardt \textit{et al.} \cite{Schardt2008}, who carefully monitored the beam energy and characterized the water equivalent thicknesses of the different materials the proton beam encountered before entering the liquid water target. Kumazaki \textit{et al.} show figures with measured depth-dose curves for beams of exactly $190.00$, $150.05$, and $230.00$ MeV. Similar measurements were performed by Schardt \textit{et al.}, but their depth-dose curves were not plotted. In the work by Paul \cite{Paul2013ACQ}, he reviews Schardt \textit{et al.}'s data, providing numerical values for the depths of the Bragg peak maxima and the penetration ranges (dose falling to $80$ \% of maximum after the peak) for beams of exact energies $125.91$, $151.09$, $176.27$, and $201.46$ MeV. We will use this experimental data to compare our simulated depth-dose curves employing different $I$-values, using exactly these beam energies in the simulations. Reviewing the numbers given in Table \ref{tab:Is}, sensible $I$-values to study are $75.0$, $78.0$, $78.5$, $79.4$, $80.0$, and $83.9$ eV.

Figure~\ref{fig:BraggHighT} shows the Bragg curves for different proton energies ranging from $125.91$ to $230.00$ MeV. Symbols show the measurements by Schardt \textit{et al.} \cite{Schardt2008} and by Kumazaki \textit{et al.} \cite{Kumazaki2007}, while lines represent simulation results using different $I$-values. In all cases, an $I$-value in the range from $78.0$ to $80.0$ eV provides results compatible with the experiments, while the simulations using $I = 75.0$ eV and $I = 83.9$ eV fall either too short or too far. 

\begin{figure}
    \centering
    \includegraphics[width=0.48\textwidth]{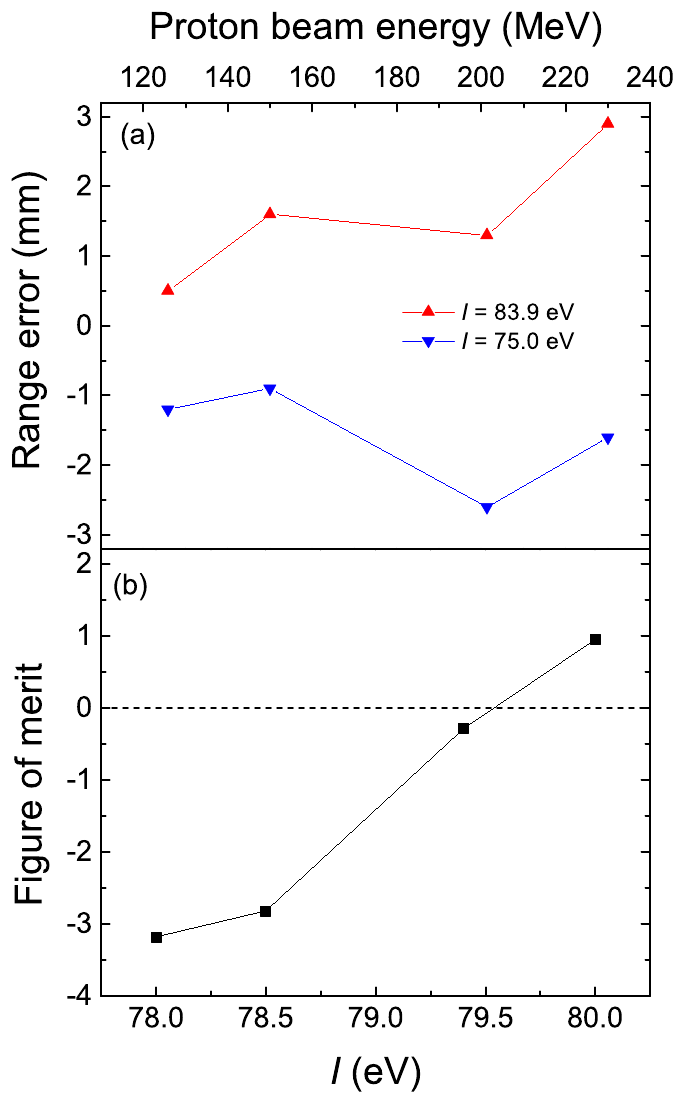}
    \caption{(a) Absolute errors for the simulated proton ranges in liquid water, with respect to experimental measurements, for the largest and lowest $I$-values used in simulations.
    (b) Figure of merit, as defined by Eq.~(\ref{eq:FoM}), obtained using different reasonable $I$-values for liquid water.}
    \label{fig:FoM}
\end{figure}

For the lowest energies studied by each group ($125.91$ MeV for Schardt \textit{et al.} and $150.05$ MeV for Kumazaki \textit{et al.}), $I = 75.0$ eV underestimates the penetration range in $-1.2$ mm ($-1.0$ \%) and $-0.9$ mm ($-0.6$ \%) respectively. On the other hand, $I = 83.9$ eV overestimates the penetration range in $+0.5$ mm ($+0.4$ \%) and $+1.6$ mm ($+1.1$ \%) respectively. These trends are shown in Fig. \ref{fig:FoM}(a).

For the highest energies ($201.46$ MeV for Schardt \textit{et al.} and $230.00$ MeV for Kumazaki \textit{et al.}), $I = 75.0$ eV underestimates the penetration range in $-2.6$ mm ($-1.0$ \%) and $-1.6$ mm ($-0.5$ \%) respectively. Regarding $I = 83.9$ eV, these simulations overestimate the penetration range in $+1.3$ mm ($+0.5$ \%) and $+2.9$ mm ($+1.0$ \%) respectively. These results are also gathered in Fig. \ref{fig:FoM}(a).

As shown in Fig.~\ref{fig:BraggHighT}, no single value of $I$ yields an optimal result across all measured Bragg curves. For all energies measured by Schardt \textit{et al.} \cite{Schardt2008},
$I = 80.0$ eV gives the best agreement, except for $176.27$ MeV, which is $I = 79.4$ eV.
For Kumazaki \textit{et al.}'s measurements, $I = 78.5$ eV gives the best result at $150.05$ and $190.00$ MeV, while at $230.00$ MeV, $I = 78.0$ eV is the best value. These differences may arise from uncertainties in determining the water-equivalent thicknesses used in each experiment for the materials traversed by the proton beam before it impinges on the liquid water target, or perhaps from uncertainties in initial beam energy or broadening. 

It would then be convenient to judge each $I$-value by comparing the cumulative errors it provides with measurements for all beam energies for which Kumazaki \textit{et al.} and Schardt \textit{et al.} provide results. For this purpose, we have evaluated, for each combination of beam energy $T_0$ and $I$-value, the relative error (in percentage) for the position $z_{{\rm max}}$ of the maximum in the dose:
\begin{equation}
    \epsilon_{{\rm max}}(T_0, I) = \frac{z_{{\rm max}}^{\rm sim}(T_0, I)-z_{{\rm max}}^{\rm exp}(T_0)}{z_{{\rm max}}^{\rm exp}(T_0)} \times 100 \, \mbox{.}
\end{equation}
Similarly, we evaluate the relative error in the range position, $\epsilon_{\rm range}$. Then, we have calculated a figure of merit (${\rm FoM}$) for each $I$-value, which we define as the sum, over all beam energies $T_0$, of all relative errors in the position of the dose maximum and in the range:
\begin{equation}
    {\rm FoM}(I) = \sum_{T_0} \left[ \epsilon_{{\rm max}}(T_0, I) + \epsilon_{{\rm range}}(T_0, I) \right]  \, \mbox{.}
    \label{eq:FoM}
\end{equation}

Figure \ref{fig:FoM}(b) depicts the results of this figure of merit for each of the plausible $I$-values studied. Negative values denote a global underestimation in the simulations of the measured depths for a particular $I$-value, while positive values denote a global overestimation. As can be seen, out of the $I$-values studied, that of $79.4$ eV arising from the MELF-GOS method (and supported by the TDDFT-Penn approach) provides the figure of merit closest to $0$, demonstrating the overall best reproduction of all the experimental Bragg curves carefully measured by Schardt \textit{et al.} and Kumazaki \textit{et al.}. Since this $I$-value comes from a theoretical approach well suited to calculating stopping powers for charged particles and is based on the most reliable experimental source for the optical properties of liquid water, it is reasonable that it yields the best simulated outcomes. Based on these results, we consider $I = 79.4$ eV to be the best current estimate of the mean excitation energy of liquid water.

\section{\label{sec:conclusions}Summary and conclusions}

The stopping power of liquid water for proton beams has been determined theoretically and semiempirically in a wide energy range. From the theoretical point of view, the dielectric formalism within the MELF-GOS method and the TDDFT-Penn approaches have been employed. 
Both methods converge for proton energies above $200$ keV, reinforcing their accuracy in the energy range more relevant for protontherapy. Importantly, the TDDFT-Penn results demonstrate that the mass stopping power of liquid water and amorphous ice is identical for all energies of practical interest in radiotherapy \cite{Matias2025}, providing a means to avoid the experimental difficulties in measuring the stopping power of liquid water.

On the other hand, the Varelas-Biersak formula proposed by ICRU Report 49 \cite{ICRU49} has been successfully fitted to the two sets of experimental determinations of the stopping power of liquid water for protons \cite{Shimizu2009,Shimizu2010,Siiskonen2011} (complemented by SRIM calculations \cite{SRIM2013} at low energies). However, this data set underestimates the theoretical calculations by around $10$\% in the intermediate energy range of $0.3-2$ MeV. 

The MELF-GOS and TDDFT-Penn approaches predict a value of $I = 79.4$ eV for the mean excitation energy of liquid water, based on the most recent experimental determination of its dielectric properties \cite{Hayashi2000}. The value recommended by ICRU Reports 49 \cite{ICRU49} and 37 \cite{ICRU37}, and adopted by NIST's PSTAR database \cite{Berger2005}, is somewhat smaller ($75$ eV), while direct fitting of the Bethe equation to Siiskonen \textit{et al.}'s high energy experimental data \cite{Siiskonen2011} gives a larger value ($83.9$ eV). The scientific literature accumulated in the last decades \cite{Bichsel2000,Kramer2000,Kumazaki2007,Schardt2008,Emfietzoglou2009,GarciaMolina2009,Paul2013ACQ,Dingfelder2014,Faddegon2015} indicates a sensible value between $78$ and $80$ eV. 

Using different stopping-power models, we first simulated the depth-dose curves for low-energy protons ($1-50$ MeV) in liquid water. The results from the MELF-GOS and TDDFT-Penn approaches are almost identical, indicating that the details of the stopping power below $200$ keV are not particularly relevant in this beam energy range. Simulations using the formula by Varelas-Biersak fitted to the experimental measurements of the stopping power of liquid water for protons \cite{Shimizu2009,Shimizu2010,Siiskonen2011,Siiskonen2011} overestimate the penetration range by $\sim 2$\% at $10$ MeV and $\sim 14$\% at $1$ MeV, as a result of the underestimation of the stopping power in the intermediate energy range. However, for larger beam energies, the differences become insignificant.

Still, it has been shown that the penetration range at the energies common in protontherapy ($100-250$ MeV) is rather sensitive to the $I$-value used in the simulations. The use of $I = 75.0$ or $83.9$ eV can lead to range errors of up to $\pm 3$ mm at some beam energies, with respect to carefully measured depth-dose curves \cite{Kumazaki2007,Schardt2008}. On the contrary, $I$-values in the range $78-80$ eV lead to very good agreement with these measured Bragg curves.

We have performed a statistical analysis of the relative errors in the position of the dose maximum and the penetration range in simulations at $I = [78-80]$ eV, with respect to the experimental Bragg curves for well-defined incident energies. The results show that the minimum cumulative errors across all analyzed beam energies are obtained from simulations with $I = 79.4$ eV. This is precisely the value predicted from the optical properties of liquid water and supported by the stopping powers obtained using both the MELF-GOS and TDDFT-Penn methods. Overall, the good comparison of these methods with experimental stopping-power determinations for protons in liquid water and amorphous ice, together with the minimization of errors in depth-dose simulations, provides solid evidence that the mean excitation energy of liquid water is very close to $79.4$ eV.



\begin{acknowledgments}
This work is part of the R\&D project no. PID2021-122866NB-I00 funded by the Spanish Ministerio de Ciencia e Innovación (MCIN/AEI/10.13039/501100011033/) and by the European Regional Development Fund (``ERDF A way
of making Europe''), as well as of the R\&D project no. 22081/PI/22 funded by the Autonomous Community of the Region of Murcia through the call ``Projects for the development of scientific and technical research by competitive groups'', included in the Regional Program for the Promotion of Scientific and Technical Research (Action Plan 2022) of the Fundación Séneca – Agencia de Ciencia y Tecnología de la Región de Murcia.
This work has also been done as part of the Project INCT-Física Nuclear e Aplicações, Projeto No. 408419/2024-5. 
\end{acknowledgments}




\bibliography{library}

\end{document}